# Evidence for GN-z11 as a luminous galaxy at redshift 10.957

Linhua Jiang[1,2], Nobunari Kashikawa[3,4], Shu Wang[1,2], Gregory Walth[5], Luis C. Ho[1,2], Zheng Cai[6], Eiichi Egami[7], Xiaohui Fan[7], Kei Ito[3,8], Yongming Liang[3,8], Daniel Schaerer[9], and Daniel P. Stark[7]

*[1]Kavli Institute for Astronomy and Astrophysics, Peking University, Beijing, China*

*[2]Department of Astronomy, School of Physics, Peking University, Beijing, China*

*[3]Department of Astronomy, Graduate School of Science, The University of Tokyo, Tokyo, Japan*

*[4]Optical and Infrared Astronomy Division, National Astronomical Observatory, Tokyo, Japan*

*[5]Observatories of the Carnegie Institution for Science, Pasadena, CA, USA*

*[6]Department of Astronomy, Tsinghua University, Beijing, China*

*[7]Steward Observatory, University of Arizona, Tucson, AZ, USA*

*[8]Department Astronomical Science, SOKENDAI (The Graduate University for Advanced Studies), Tokyo, Japan*

*[9]Geneva Observatory, University of Geneva, Geneva, Switzerland*

**GN-z11 was photometrically selected as a luminous star-forming galaxy candidate at redshift $z > 10$ based on Hubble Space Telescope (HST) imaging data[1]. Follow-up HST near-infrared grism observations detected a continuum break that was explained as the Lyα break corresponding to $z = 11.09^{+0.08}_{-0.12}$ (ref. 2). However, its accurate redshift remained unclear. Here we report a probable detection of three ultraviolet (UV) emission lines from GN-z11, which can be interpreted as the [C III] λ1907, C III] λ1909 doublet and O III] λ1666 at $z = 10.957 \pm 0.001$ (when the Universe was only ~420 Myr old, or ~3% of its current age). This is consistent with the redshift of the previous grism observations, supporting GN-z11 as the most distant galaxy known to date. Its UV lines likely originate from dense ionized gas that is rarely seen at low redshifts, and its strong [C III] and C III] emission is partly due to an active galactic nucleus (AGN) or enhanced carbon abundance. GN-z11 is luminous and young, yet moderately massive, implying a rapid build-up of stellar mass in the past. Future facilities will be able to find the progenitors of such galaxies at higher redshift and probe the cosmic epoch in the beginning of re-ionization.**

Owing to recent advances of instrumentation on the HST and large ground-based telescopes, a number of high-redshift galaxies have been discovered during the past decade[3-7]. They provide deep insights on the first several hundred million years (Myr) of cosmic history. The majority of the currently known galaxies at $z \geq 7.5$ were detected by the HST Wide Field Camera 3 (WFC3). Meanwhile, a small fraction of them at redshifts up to $z \sim 9$ have been spectroscopically confirmed[8-12], primarily from the Lyα emission line and far-infrared fine-structure lines. GN-z11 has an apparent magnitude of 26.0 in the HST WFC3 F160W band (magnitudes are on the AB system). Its brightness makes ground-based observations feasible. In 2017 and 2018, we performed near-infrared spectroscopic observations using the multi-object spectrograph MOSFIRE[13] on the Keck I telescope. The total on-source integration time in the $H$ and $K$ bands are 4.3 and 5.3 hours, respectively; see Methods for details.

Figure 1 and Table 1 summarize the Keck spectroscopy and our observational results. We first verify the detection of the UV continuum emission from GN-z11, which is too faint to directly appear in the two-dimensional (2D) spectroscopic images. We stack the 2D $K$-band spectrum along the wavelength direction and detect a signal with a 5.1σ significance at the expected spatial position, indicative of its origin from the GN-z11 UV continuum (Fig. 1). We then search for line emission in the $K$ band and identify a line at 19922 Å and a line pair at 22797 and 22823 Å, with significance of 3.3σ, 2.6σ, and 5.3σ, respectively. Taking the broadband photometry into consideration, we rule out the possibility that these lines are from a low-redshift galaxy (Methods). If we disregard two weak detections of 3.3σ and 2.6σ, the strongest line with the 5.3σ detection can be explained as [C III] λ1907 at $z = 10.970$ or C III] λ1909 at $z = 10.957$. Alternatively, the three lines can be interpreted as O III] λ1666 and the [C III] λ1907, C III] λ1909 doublet at $z = 10.957$. We favor this interpretation, as the expected C III] λ1909 emission is not detected if the 5.3σ line is [C III] λ1907 (Methods). This redshift agrees with the redshift from previous HST low-resolution grism observations[2].

Figure 2 shows the 2D and one-dimensional (1D) spectra of the [C III] λ1907 and C III] λ1909 doublet. This doublet is commonly seen in line-emitting galaxies at redshifts up to $z > 7$ (refs. 14,15). C III] λ1909 is the strongest emission line that we detect from GN-z11. We fit a Gaussian profile to its 1D spectrum and calculate redshift and line flux from the best-fitting parameters. The resultant redshift is $z = 10.957 \pm 0.001$. We adopt this redshift as the systemic redshift. The C III] λ1909 line flux is $(3.5 \pm 0.7) \times 10^{-18}$ erg s$^{-1}$ cm$^{-2}$, and the rest-frame equivalent width (EW$_0$) is $28 \pm 5$ Å. The [C III] λ1907 line is weaker, with a flux of $(1.5 \pm 0.6) \times 10^{-18}$ erg s$^{-1}$ cm$^{-2}$ and an EW$_0 = 12 \pm 5$ Å. The total EW$_0$ of the doublet in GN-z11 is high and comparable to the largest values reported in lower-redshift galaxies[16-18] (Fig. 3). The full width at half maximum (FWHM) of the C III] λ1909 line, corrected for instrumental resolution, is ~$92 \pm 23$ km s$^{-1}$. We estimate a lower limit for the dynamic mass of ~$2 \times 10^9$ $M_\odot$ ($M_\odot$ is the solar mass), which is roughly 1.5 times the stellar mass that will be derived below (Methods).

The forbidden transition [C III] $\lambda$1907 and the intercombination transition C III] $\lambda$1909 have very different critical densities for collisional de-excitation, so their line flux ratio traces the electron density. In typical nebular environments with electron densities $\leq 10^4$ cm$^{-3}$, the [C III] $\lambda$1907 to C III] $\lambda$1909 flux ratio is roughly between 1.0 and 1.6. This ratio decreases rapidly toward higher densities, but nebular environments with densities higher than $10^4$ cm$^{-3}$ are rare[16,19]. We observe [C III] $\lambda$1907/C III] $\lambda$1909 = 0.43 ± 0.18, which implies a density of $(4-13) \times 10^4$ cm$^{-3}$ (1σ range). This suggests that the C III] doublet emission in GN-z11 arises from a very dense interstellar medium (ISM). A high electron density is favored by theory[17] to explain the large C III] $EW_0$ in GN-z11.

The O III] $\lambda$1666 line has a flux of $(1.7 \pm 0.5) \times 10^{-18}$ erg s$^{-1}$ cm$^{-2}$ and an $EW_0$ of 10 ± 3 Å. This line usually forms a doublet with the much weaker O III] $\lambda$1661 line in line-emitting galaxies[20-22]. O III] $\lambda$1661 is not detected in GN-z11. We have searched for He II $\lambda$1640 in the $K$ band, but unfortunately its wavelength position is severely contaminated by strong OH skylines. We do not detect any significant emission signal in the $H$ band, which covers a rest-frame wavelength range of 1230−1500 Å. The other two potentially strong lines, Lyα (if not completely absorbed by the neutral intergalactic medium) and C IV $\lambda$1549, are located in two water vapor absorption bands in the near-infrared, and thus are not reachable from the ground.

We use photoionization models[17,23] to explore the UV line emission from GN-z11 (Fig. 3 and Methods). The C III] doublet is often the strongest emission feature after Lyα in the wavelength range that we cover[25]. The typical $EW_0$ of the doublet is about a few Å, which can be explained by normal stellar populations (black curve in Fig. 3a). However, the origin of strong C III] emission with $EW_0 \geq 20$ Å is not fully understood. Possible mechanisms responsible for such strong emission include AGN, rare and massive stellar populations, and enhanced carbon abundances[17]. In addition, a high ISM density, as present in our case, also helps. Fig. 3a shows that the C III] emission strength (C III] stands for the combination of the [C III], C III] doublet here) is nearly doubled if an AGN contributes 10% of ionizing photons (blue curve). The C III] $EW_0$ is further doubled and reaches the value of GN-z11 if the carbon abundance is higher than normal (orange curve). Alternatively, the C III] emission arising entirely from an AGN narrow-line region (NLR) can easily reach a very high $EW_0$ (purple curves). The narrow C III] line width does not suggest the presence of a type 1 AGN, but the available X-ray data are not deep enough to rule out weak or type 2 nuclear activity in GN-z11.

Fig. 3a suggests that in GN-z11 an AGN may contribute a non-negligible fraction of ionizing photons and the carbon abundance may also be enhanced. The C III] to O III] $\lambda$1666 line flux ratio shown in Fig. 3b is broadly consistent with those of lower redshift galaxies, where this ratio spans a large range from < 2 to > 6. Strong AGN radiation tends to suppress O III] and thus increase C III]/O III], but this line ratio alone is not sufficient to distinguish different mechanisms (e.g., type 2 AGN versus enhanced carbon abundance). Future observations of more diagnostic emission lines, such as C IV and He II, are needed to draw firmer conclusions[17,22,23].

We measure the UV continuum slope β ($f_\lambda \propto \lambda^\beta$, where $f$ is flux density and λ is wavelength) using the redshift 10.957 and the broadband photometry in the WFC3 F140W, F160W, and Spitzer IRAC channel 1 (3.5 μm) bands. The resultant slope β = −2.4 ± 0.2 is similar to a previous measurement[2] and indicates little or no dust reddening. Assuming no dust reddening, the star formation rate (SFR) calculated from the UV continuum flux is 26 ± 3 $M_\odot$ yr$^{-1}$ (assuming $H_0$ = 68 km s$^{-1}$ Mpc$^{-1}$, $\Omega_m$ = 0.3, and $\Omega_\Lambda$ = 0.7). We model the broadband spectral energy distribution (SED) and constrain three physical quantities of the stellar population: dust reddening, stellar mass, and age (Methods and Table 1). The best-fit dust reddening is consistent with zero. While age is usually poorly constrained by SED modeling of a few photometric data points, our result suggests a young stellar population with an age 70 ± 40 Myr. This requires a rapid build-up of its stellar mass, $(1.3 \pm 0.6) \times 10^9$ $M_\odot$.

We have used a single stellar population to model the broadband SED. Recent studies have shown that two populations are common at $z > 6$, an evolved population and a recent starburst likely induced by galaxy merging[12,14,26,27]. The same may hold for GN-z11: its moderate stellar mass suggests that a relatively evolved population must be present, while a very young (a few Myr) population would be needed to explain the high C III] EW$_0$ if AGN contribution is not important. This has small impact on our measurement of stellar mass, but the assumption of a single age would be inappropriate. The specific SFR (~2.0 × 10$^{-8}$ yr$^{-1}$), the ratio of SFR to stellar mass, is similar to that of luminous star-forming galaxies at $z \sim 6$-7 (ref. 28). Although luminous and massive galaxies like GN-z11 are rare at $z \sim 11$, similar galaxies have been identified in cosmological simulations[29,30].

Multiple lines of evidence indicate that our line detection corresponding to redshift 10.957 for GN-z11 is reliable. Further confirmation of the three UV lines and the detection of additional lines are certainly desirable, as would deeper observations that enable fuller characterization of the source's physical properties. Despite its youth, GN-z11 has already accumulated substantial stellar mass, and observations of even higher redshifts are needed to probe the earliest phases of the formation and evolution of the first galaxies.

Correspondence and requests for materials should be addressed to L.J. (jiangKIAA@pku.edu.cn) or N.K. (n.kashikawa@astron.s.u-tokyo.ac.jp).

**Acknowledgements** We acknowledge support from the National Science Foundation of China (11721303, 11890693, 11991052), the National Key R&D Program of China (2016YFA0400702, 2016YFA0400703), and the Chinese Academy of Sciences (CAS) through a China-Chile Joint Research Fund (1503) administered by the CAS South America Center for Astronomy. N.K. acknowledges support from the JSPS grant 15H03645. We thank P. Oesch and C. Steidel for helpful discussions on observations and data reduction. We thank A. Feltre, K. Nakajima, and T. Nanayakkara for providing data shown in Fig. 3. The data presented herein were obtained at the W. M. Keck Observatory, which is operated as a scientific partnership among the California Institute of Technology, the University of California, and the National Aeronautics and Space Administration. The Observatory was made possible by the generous financial support of the W. M. Keck Foundation. We wish to recognize and acknowledge the very significant cultural role and reverence that the summit of Mauna Kea has always had within the indigenous Hawaiian community. We are most fortunate to have the opportunity to conduct observations from this mountain.

**Author contributions** L.J. designed the program, carried out the Keck observations, analysed the data, and prepared the manuscript. N.K. designed the program and carried out the observations. S.W. and G.W. reduced the images. L.H. helped to prepare the manuscript. K.I. and Y.L. assisted with the observations. All authors helped with the scientific interpretations and commented on the manuscript.

**Competing interests** The authors declare that they have no competing interests.

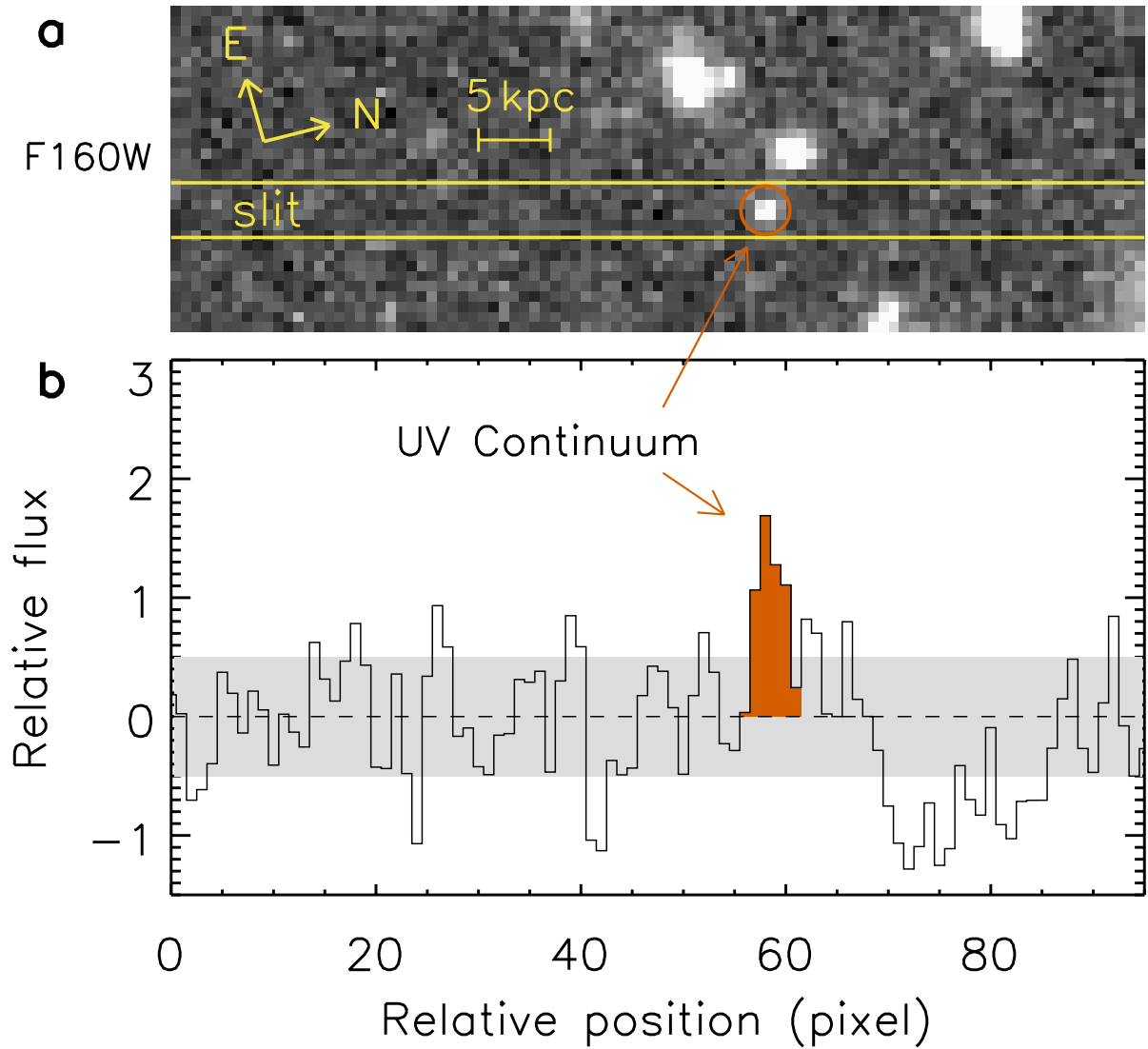

**Fig. 1 Keck MOSFIRE spectrum of GN-z11.** (a) Demonstration of our observations. The background image is part of the HST WFC3 F160W image. It has been rotated and re-binned so that the pixel scale is the same as that of the MOSFIRE image scale of ~0.18" per pixel. The image orientation and physical scale are shown in yellow notation. The two lines represent the 0.9"-wide slit. The UV continuum emission is enclosed by the circle. (b) Detection of the GN-z11 continuum. The spectrum shows the average flux stacked along the wavelength direction of the 2D *K*-band spectrum, as a function of spatial position (along the slit). The grey area represents the 1σ uncertainty region. A signal (in vermillion) was detected at a significance of 5.1σ, with a position coincident with GN-z11 in the F160W image, suggesting its origin from the GN-z11 continuum emission.

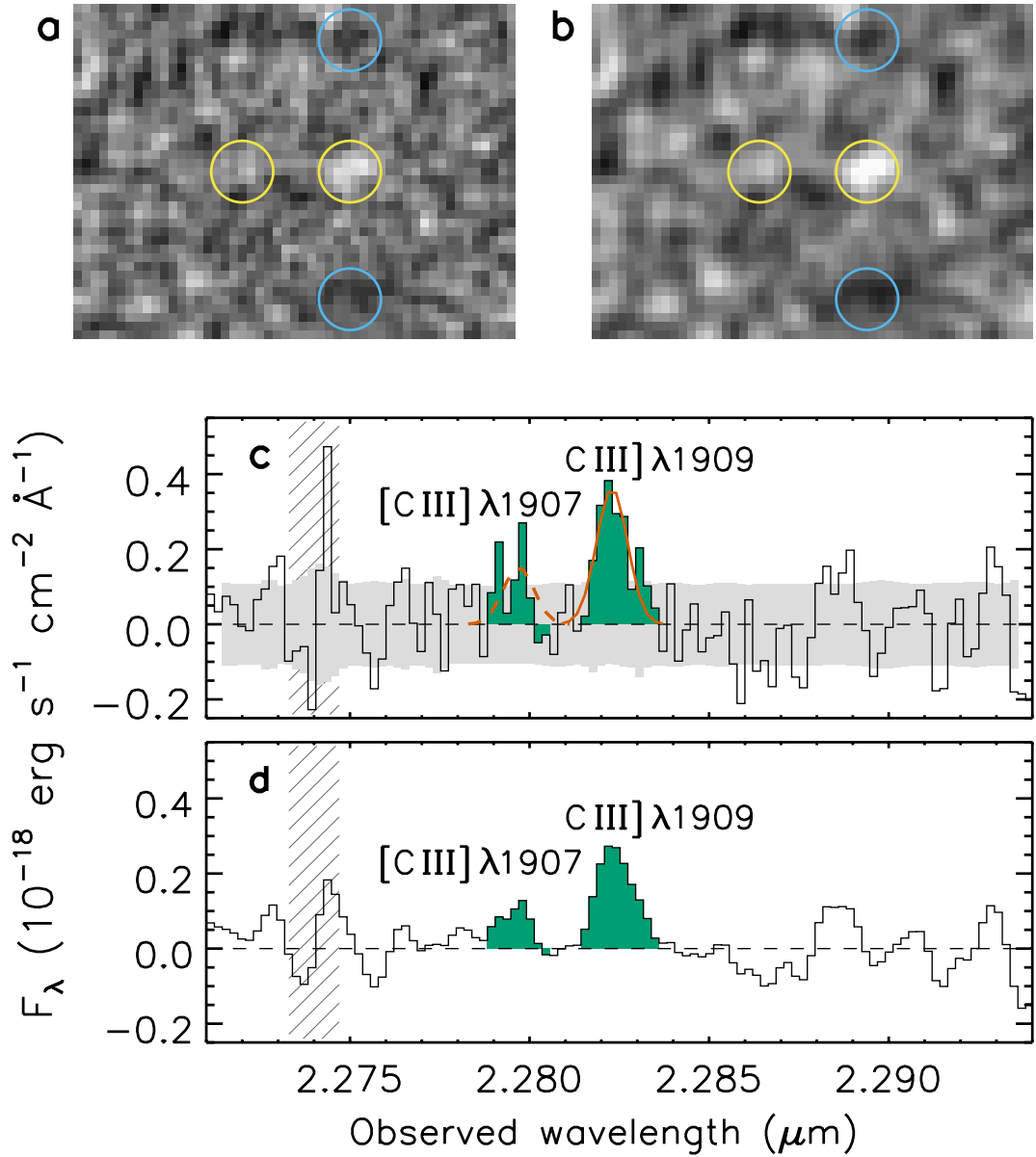

**Fig. 2 Detection of the [C III] λ1907 and C III] λ1909 emission lines.** (a) Part of the *K*-band 2D spectrum with the two lines (enclosed by the vermillion circles) detected at a significance of 2.6σ and 5.3σ, respectively. The two negative signals enclosed by the blue circles are produced by our data reduction for standard ABBA observations (Methods). (b) A smoothed version of the 2D spectrum to better illustrate the detection of the lines; a Gaussian filter with a size of 2.5 pixels is used. (c) Extracted 1D spectrum. The grey area represents the 1σ uncertainty region. The hatched area indicates a region affected by skylines. The two emission lines are shown in green. The vermillion solid profile is the best-fit Gaussian to the C III] λ1909 line. The vermillion dashed profile is the same Gaussian profile, but scaled to match the [C III] λ1907 flux and shifted to match the [C III] λ1907 wavelength. (d) Same as (c), but of the smoothed 2D spectrum (b).

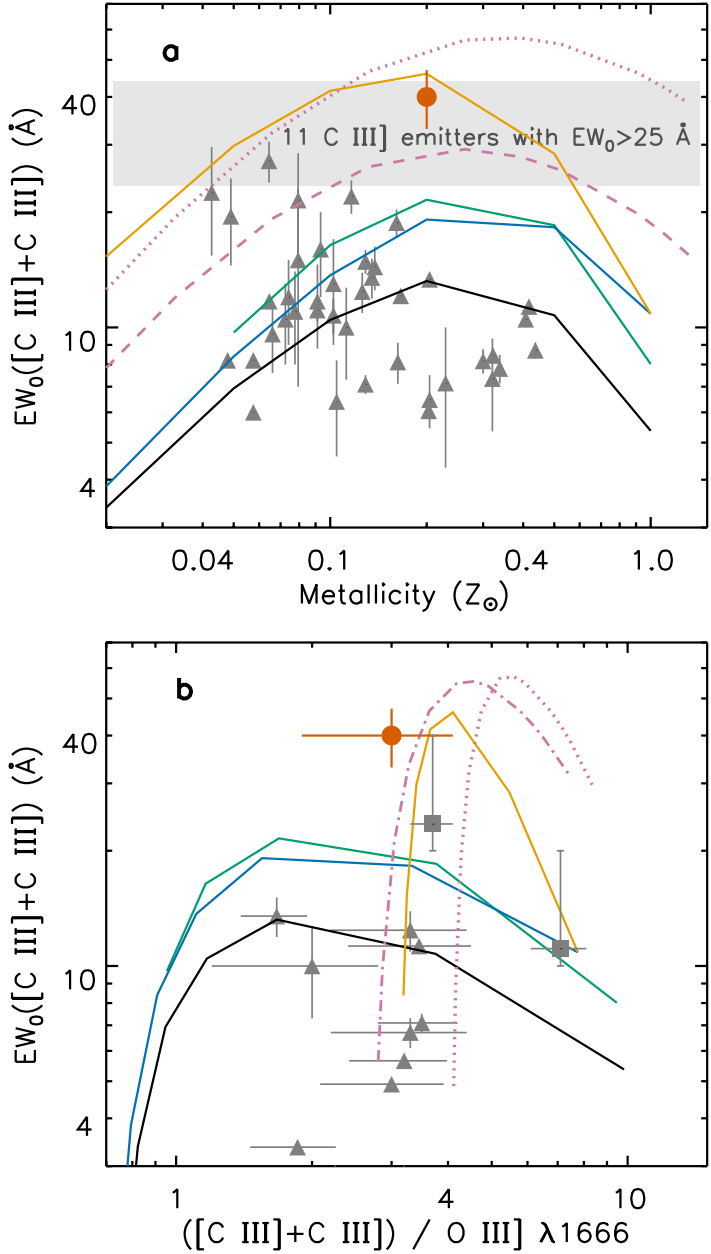

**Fig. 3 Comparison of observations with photoionization models**. (a) Comparison with lower redshift galaxies having the strongest C III] emission known in the literature. The triangles represent a collection[17] of strong C III] emitters with $EW_0 > 6$ Å. The grey area represents an approximate locus occupied by a sample[17,18] of 11 extreme C III] emitters with $EW_0 > 25$ Å that are believed to be partly powered by AGN. Their metallicity information is not available. The vermillion circle is GN-z11, whose metallicity is assumed to be 0.2 $Z_\odot$ ($Z_\odot$ is the solar metallicity). The error bars indicate 1σ uncertainties. The curves represent a single stellar population model (Model 1; black), a binary stellar population model (Model 2; green), Model 1 plus 10% contribution from an AGN (Model 3; blue), Model 3 plus an enhanced carbon abundance (Model 4; orange), and two AGN NLR models (Models 5 and 6; dotted and dashed purple). (b) Model prediction for the ([C III]+C III])/O III] line ratio. The triangles

represent a collection of three lower redshift galaxy samples[21,22,24]. The two squares represent the average line ratio of two samples[18] with 10 Å < C III] $EW_0$ < 20 Å and 20 Å < C III] $EW_0$ < 40 Å. The vertical error bars for the squares indicate the $EW_0$ ranges that the two samples cover. The coloured curves are the same as those for Models 1-5 in (a), except for the dash-dotted purple curve, which is a slightly modified version of Model 5 for the AGN NLR. This analysis suggests that an AGN may contribute a non-negligible fraction of ionizing photons in GN-z11, and that carbon may also be enhanced (see Methods for more details).

Table 1. Properties of GN-z11

| Parameters | Values |
|---|---|
| R.A. (hh:mm:ss) | 12:36:25.46 |
| Dec. (º:':") | +62:14:31.4 |
| Redshift | $10.957 \pm 0.001$ |
| Flux ([C III] λ1907) (erg s$^{-1}$ cm$^{-2}$) | $(1.5 \pm 0.6) \times 10^{-18}$ |
| $EW_0$ ([C III] λ1907) (Å) | $12 \pm 5$ |
| Flux (C III] λ1909) (erg s$^{-1}$ cm$^{-2}$) | $(3.5 \pm 0.7) \times 10^{-18}$ |
| $EW_0$ (C III] λ1909) (Å) | $28 \pm 5$ |
| FWHM (C III] λ1909) (km s$^{-1}$) | $92 \pm 23$ |
| Flux (O III] λ1666) (erg s$^{-1}$ cm$^{-2}$) | $(1.7 \pm 0.5) \times 10^{-18}$ |
| $EW_0$ (O III] λ1666) (Å) | $10 \pm 3$ |
| UV continuum slope β | $-2.4 \pm 0.2$ |
| UV SFR ($M_\odot$ yr$^{-1}$) | $26 \pm 3$ |
| Specific SFR (yr$^{-1}$) | $(2.0 \pm 0.9) \times 10^{-8}$ |
| Dust reddening $E(B-V)$ (mag) | $0.01 \pm 0.01$ |
| Age (Myr) | $70 \pm 40$ |
| Stellar mass ($M_\odot$) | $(1.3 \pm 0.6) \times 10^9$ |

## Methods

**Keck observations and data reduction.** We carried out *H*- and *K*-band spectroscopic observations of GN-z11 using the near-infrared, multi-object spectrograph MOSFIRE[13] on the Keck I telescope. Our main purpose was to detect emission lines such as C III] λ1909, [C III] λ1907, O III] λ1666, and He II λ1640 that exist in a significant fraction of galaxies at $z \geq 2$ (refs. 25,31). MOSFIRE has an effective field-of-view of 3 by 6 arcminutes. Our slit mask was designed to place GN-z11 near the center of the field, with a position angle (345.5˚) that minimizes light pollution from nearby objects (Fig. 1a). One bright reference star was chosen for a slit next to GN-z11, and it was used for flux calibration. The slit mask also includes 18 targets at lower redshifts.

The *K*-band spectroscopy was made under good observing conditions on 2017 April 7 (universal time), and the seeing ranged from 0.6" to 0.7". We obtained 5.3 hours of on-source integration. The individual exposure time was 179 seconds. The *H*-band spectroscopy was made on 2018 March 7 and 8. The observing conditions were moderate to poor, with passing clouds and varying seeing between 0.8" and 1.2". The individual exposure time was 120 seconds, and a total of 4.3 hour on-source integration was obtained. We used the classic ABBA observing mode. The slit width was 0.9", which delivers a resolving power of ~2800. We also observed a spectrophotometric standard star HIP 68767 (spectral type A0) prior to or after the science observations.

The MOSFIRE images were reduced using a publicly available data reduction pipeline (https://github.com/Keck-DataReductionPipelines/MosfireDRP). The pipeline performs all necessary reduction steps, including flat-fielding, sky subtraction, wavelength calibration, rectification, and co-addition. We followed the basic reduction procedure and made a few minor changes. The dither information in the image headers provides an initial guess of target positions. It has been known that there is a small and systematic drift of target positions in MOSFIRE data[32]. We calculated this drift by measuring the position of the reference star in each image, and corrected the drift by shifting individual images. Another change to the pipeline was to identify cosmic rays and interpolate over them before we fed the images to the pipeline. Although the pipeline does sigma-clipping during image stacking, a pre-rejection of cosmic rays reduces the effect from re-sampling pixels with cosmic rays. Note that when we combined the *K*-band images, we excluded an image that contained a bright flash signal. The final output products consist of fully reduced 2D spectra and signal-to-noise ratio images, as well as other ancillary data.

The data were reduced by multiple members in our group, and the results are consistent. In addition, one of the MOSFIRE team members helped us process the images independently and obtained consistent results for the detection of the C III] and O III] lines. The ABBA observing mode produces a standard negative-positive-negative pattern in the 2D combined spectrum. This pattern is clearly seen for C III] λ1909 and O III] λ1666. The [C III] λ1907 line is not prominent in either the A or B position, because it is only a 2.6σ detection in the combined spectrum. In addition, we evenly split the K-band data and produced two 2D spectra. The C III] λ1909 emission line is detected at ~3.5σ and clearly seen in either spectrum (Extended Data Figs. 1 and 2).

**Detection of emission lines.** We first verify the detection of the UV continuum emission by stacking the 2D K-band spectrum along the wavelength direction. We detect a signal with a 5.1σ significance at the expected spatial position of the GN-z11 UV continuum (Fig. 1). We also see the standard negative-positive-negative pattern in Fig. 1b. In our ABBA observing mode, the separation between the A and B positions was 3", or ~16.7 pixels. The peak of the positive signal is roughly at $x \sim 58$ in Fig. 1b, so we expect to see two negative signals at $x \sim$ 41 and 75, respectively. The negative signal at $x \sim 41$ is clearly seen. We can also see the negative signal at $x \sim 75$, although it is in a big trough that makes it less obvious.

We search for emission lines in the K-band 2D spectrum and first identify a strong (5.3σ significance) line emission feature at about 22823 Å. Meanwhile, we detect a weaker (2.6σ significance), nearby line at 22797 Å. This pair of lines can be explained as the [C III] λ1907, C III] λ1909 doublet at $z = 10.957$. We would not have claimed a 2.6σ line as a detection if this line does not form a [C III], C III] doublet that is commonly seen at high redshift. We then search for >3σ lines that are associated with this redshift, and detect a line (3.3σ) at ~19922 Å that is consistent with O III] λ1666 (Extended Data Fig. 3). We do not detect any other lines in the spectrum at greater than 3σ significance. If the two weak detections of 3.3σ and 2.6σ are not considered, the strongest line with the 5.3σ detection can be explained as [C III] λ1907 at $z = 10.970$ or C III] λ1909 at $z = 10.957$. If this line is [C III] λ1907 at $z =$ 10.970, we would expect to detect C III] λ1909 with significance of $\geq 3\sigma$, because the largest flux ratio of [C III] λ1907 to C III] λ1909 is about 1.6 in regular environments. Since we did not detect the expected C III] λ1909 emission, the 5.3σ line is not likely [C III] λ1907. Therefore, we interpret the line pair at 22797 and 22823 Å as the [C III] λ1907, C III] λ1909 doublet and the line at 19922 Å as O III] λ1666 at $z = 10.957$.

The spatial location of the above emission lines and the collapsed continuum signal (Fig. 1b) are the same. This location is roughly consistent with the expected continuum location based on the WCS in the image header, with a marginal offset of ~2 ± 1 pixels or 1.6 ± 0.8 kiloparsecs. In order to identify a possible offset, we make use of 4 of the 5 alignment stars (one of them is slightly elongated and thus not used), and translate the WCS from R.A. and Dec. to the pixel coordinates without using the WCS in the image header. We find that the offset is reduced to $1 \pm 1$ pixel. Since these offsets are marginal, we do not try to explore their physical meaning.

There are two types of low-redshift galaxies that may contaminate a high-redshift galaxy sample: extreme line-emitting galaxies whose prominent line emission mimics strong breaks in the broadband photometry, and old/dusty galaxies with strong Balmer breaks and/or very red colors. Based on our Keck spectra and the HST broadband photometry, we can rule out these two possibilities using the line at 22823 Å, without using the other two lines. This line is the strongest ($5.3\sigma$) feature in our *H*- and *K*-band spectra. The strongest emission lines after Ly$\alpha$ in star-forming galaxies are H$\alpha$, H$\beta$, [O III] $\lambda$5007, and [O II] $\lambda$3727. If this line were H$\alpha$ at $z \sim 2.5$, H$\beta$ at $z \sim 3.7$, [O III] $\lambda$5007 $z \sim 3.6$, or [O II] $\lambda$3727 $z \sim 5.1$, then its $EW_0$ would be 96, 71, 73, or 55 Å, respectively. This would imply that GN-z11 is not an extreme line-emitting galaxy, as its broadband photometry would be completely dominated by continuum emission. Using four broadbands (WFC3 F140W and F160W, and Spitzer IRAC channels 1 and 2), we obtain a steep continuum slope of $\beta = -2.1$. We have assumed that the Balmer break (if there is one) is not located in the F140W band; otherwise, the slope would be even steeper. This clearly indicates that GN-z11 is not a dusty galaxy. Similarly, this is not a galaxy at $z \sim 3.7$, 3.6, or 5.1 with a strong Balmer break because of the blue continuum. If this line were H$\alpha$ at $z \sim 2.5$, the Balmer break would be at ~12680 Å. However, GN-z11 is not detected in the deep HST WFC3 F125W band, making it impossible to explain a steep slope with a break at ~12680 Å.

We can also use SED modeling to rule out the possibility that our detected emission lines are from a low-redshift galaxy[2]. GN-z11 was not detected in optical bands, so here we only include infrared bands: HST WFC3 F105W, F125W, F140W, and F160W bands, the ground-based *K* band, and Spitzer IRAC channels 1 and 2. The photometric values are taken from the literature[2]. The details of the SED modeling is described in section 'SED modeling' below, and the results of two examples are shown in Extended Data Fig. 4. In Extended Data Fig. 4a, we assume that the emission line at 22823 Å is [O III] $\lambda$5007, and thus we fix redshift $z =$ 3.558. The best-fit model is inconsistent with the observed photometry (reduced $\chi^2 = 10.5$). For comparison, we also show the best-fit model assuming $z = 10.957$, which is well consistent with the observed photometry (reduced $\chi^2 = 0.5$). In Extended Data Fig. 4b, we assume that the emission line at 22823 Å is [O II] $\lambda$3727, and we fix redshift $z = 5.124$. The result is similar to that shown above.

We extract 1D spectra using a boxcar with an aperture size of 5 pixels, equivalent to the slit width of 0.9". We use the spectra of the spectrophotometric standard star to compute the system throughput curve and correct for telluric absorption. Flux calibration (including aperture correction) is conducted using the bright reference star mentioned earlier. The spectrum of the reference star was scaled to match its corresponding *K*-band photometry, thereby placing it on an absolute flux scale. This means that light loss, including slit loss, has been considered. When we estimate the detection significance of the emission lines, we use the central 7 pixels (corresponding to 1.5 times the FWHM of C III] λ1909), i.e., based on an area of 7 pixels (wavelength direction = 1.5 × FWHM) by 5 pixels (spatial direction = slit width) in the 2D spectrum. We find that [C III] λ1907, C III] λ1909, and O III] λ1666 are detected at significance of 2.6σ, 5.3σ, and 3.3σ, respectively. The line fluxes of C III] λ1909 and O III] λ1666 are calculated from the best-fitting Gaussian profiles. The [C III] λ1907 flux is calculated from the C III] λ1909 flux using their flux ratio within the central 7 by 5 pixels mentioned above. The line EWs are computed using the continuum measured from the redshift 10.957 and the broadband photometry (see the main text).

We estimate the dynamic mass $M_{dyn}$ based on the C III] λ1909 line width. We assume an ideal case of a spherical and uniform density and use $M_{dyn} = 1.2 \times 10^{10} \sigma^2$ r $M_\odot$ (refs. 33,34), where σ is the velocity dispersion in units of 100 km s$^{-1}$ and r is the galaxy size in units of kpc. GN-z11 is extended over 0.6" (~2.4 kpc) in the HST WFC3 F160W band, and we adopt r = 1 kpc. The velocity dispersion σ is simply FWHM/2.35 = 40 km s$^{-1}$. The resultant mass is ~2 × $10^9$ $M_\odot$. This calculation is associated with large uncertainties, because we do not know the structure and kinematics of this galaxy, its inclination angle (if a disk), or the size of the C III]-emitting region. In addition, C$^{+2}$ has a high ionization potential, so it is likely that the C III] line width only reflects the central part of the galaxy. This means that the estimated $M_{dyn}$ above is a lower limit of the dynamic mass. This lower limit is roughly 1.5 times the stellar mass derived from the SED modeling below.

We tried to search for He II λ1640 but were unsuccessful because its wavelength position in the spectrum is severely affected by much stronger OH skylines. Future deeper and higher-resolution observations or space observations are needed to confirm this line. We do not detect strong emission lines in the *H* band. The *H*-band images were taken in relatively poor observing conditions. In addition, the *H*-band spectrum is largely affected by a number of strong OH lines. The line depth in clean regions is roughly (2.5-3.0) × 10$^{-18}$ erg s$^{-1}$ cm$^{-2}$ (3σ), shallower than in the *K*-band spectrum. Furthermore, there are no strong emission lines in this wavelength range for a galaxy at *z* = 10.957, so we do not expect to detect strong lines in the *H* band.

**Comparison with photoionization models.** Figure 3 compares GN-z11 with lower redshift star-forming galaxies. The triangles represent a collection of several samples of strong C III]-emitters from the literature[17], and we only show those with EW$_0$ > 6 Å. The grey area denotes a sample[17,18] of 11 extreme C III]-emitters with EW$_0$ > 25 Å. Their metallicity information is not available. These sources are very rare, representing ~1% of star-forming galaxies at 2 < *z* < 4 (ref. 18). The C III] emission from these galaxies is believed to be partly due to AGN[17,18].

The C III] EW$_0$ of GN-z11 (red circle) is comparable to the highest EW$_0$ values in lower redshift galaxies. The C III] EW$_0$ stands for the EW$_0$ of the [C III] $\lambda$1907, C III] $\lambda$1909 doublet here.

We use photoionization models in the literature[17,23] to explore the UV line emission in GN-z11. We consider models of star-forming galaxies[17], star-forming galaxies with AGN activity[17], and pure AGN NLRs[23]. These models use a range of metallicities from ~ 0.01 $Z_\odot$ to > 1 $Z_\odot$, depending on different models. We assume a high gas density of $10^4$ cm$^{-3}$ and an ionization parameter log $U = -2$. In Fig. 3a, the black and green curves represent a single stellar population model ('PopStar'; Model 1) and a binary stellar population model ('BPASS'; Model 2), respectively. The blue curve represents a single stellar population model with 10% contribution from an AGN (Model 3). In Models 1-3, a standard carbon abundance (or C/O ratio) is used at each metallicity. The orange curve represents Model 3 but with an enhanced carbon abundance (fixed at the solar value; Model 4). The two purple curves represent AGN NLR models[23] with a typical dust-to-metal mass ratio ~ 0.3: Models 5 (dotted curve) uses a UV ($\lambda$ < 2500 Å) continuum slope $\alpha = -1.4$ ($f_\nu \propto \nu^\alpha$, where $\nu$ is frequency), and Model 6 (dashed curve) uses $\alpha = -1.7$. AGN continuum attenuation is not considered, so the C III] EW$_0$ values from the Models 5 and 6 are lower limits.

Fig. 3b shows the C III] to O III] $\lambda$1666 line flux ratio of GN-z11, compared with three samples lower redshift galaxies (triangles)[21,22,24]. The two squares represent the average C III]/ O III] of a sample of galaxies with 10 Å < C III] EW$_0$ < 20 Å and another with 20 Å < C III] EW$_0$ < 40 Å. The vertical error bars for the squares indicate the EW$_0$ ranges that the two samples cover. C III]/O III] spans a large range in low-redshift galaxies, from < 2 to > 6, and GN-z11 falls within this range. Models 1-5 are identical to those in Fig. 3a. The dash-dotted purple curve represents an AGN NLR model that is similar to Model 5, except that log $U = -1$ instead of $-2$. AGN radiation tends to suppress O III] and thus increase C III]/O III]. However, this line ratio alone is not sufficient to distinguish between a pure AGN origin versus an enhanced carbon abundance. Future observations of more diagnostic emission lines (e.g., C IV and He II) are needed.

**SED modeling.** We fit the broadband SED using the CIGALE code[35]. We include five bands: HST WFC3 F140W, and F160W bands, the ground-based $K$ band, and Spitzer IRAC channels 1 and 2. The photometric values are taken from the literature[2]. The basic procedure is similar to those in the literature. Given the limited number of available photometric data points, the quality of the SED fitting is dominated by data quality instead of the quality of synthesis models. We use as few free parameters as possible in our SED fitting. Our spectroscopic redshift removes one critical free parameter. We adopt the Chabrier initial mass function with a mass range of 0.1-100 $M_\odot$. Constant SFR is assumed. We also fix the metallicity to 0.2 $Z_\odot$, a typical value in high-redshift galaxies[36]. The metallicity assumption has a small effect on our measurements of the quantities below. In this paper we assume that the stellar and gas-phase metallicities are the same. We then include gaseous or nebular emission (both continuum and line emission). The inclusion of nebular emission is not important here, because our wavelength range does not cover very strong lines (e.g., Hα or [O III] λ5007) that may affect the broadband SED. From SED modeling, we mainly constrain three quantities: dust reddening $E(B-V)$, age, and stellar mass. The results are shown in Table 1 (see also Extended Data Fig. 4). The resultant dust reddening is consistent with zero. We estimate SFR = $26 \pm 3$ $M_\odot$ yr$^{-1}$ from the UV continuum luminosity[37,38], where the uncertainty only includes the measurement error of the luminosity. The age, $70 \pm 40$ Myr, is poorly constrained due to the limited number of data points and various degeneracies among parameters.

**Data availability** The Keck MOSFIRE data of this work are publicly available from the Keck Observatory Archive (https://www2.keck.hawaii.edu/koa/public/koa.php). The source data for the figures within this paper are provided as Source Data files. Other data of this study are available from the corresponding authors upon reasonable request.

**Code availability** The Keck MOSFIRE data were reduced using a publicly available data reduction pipeline (https://github.com/Keck-DataReductionPipelines/MosfireDRP).

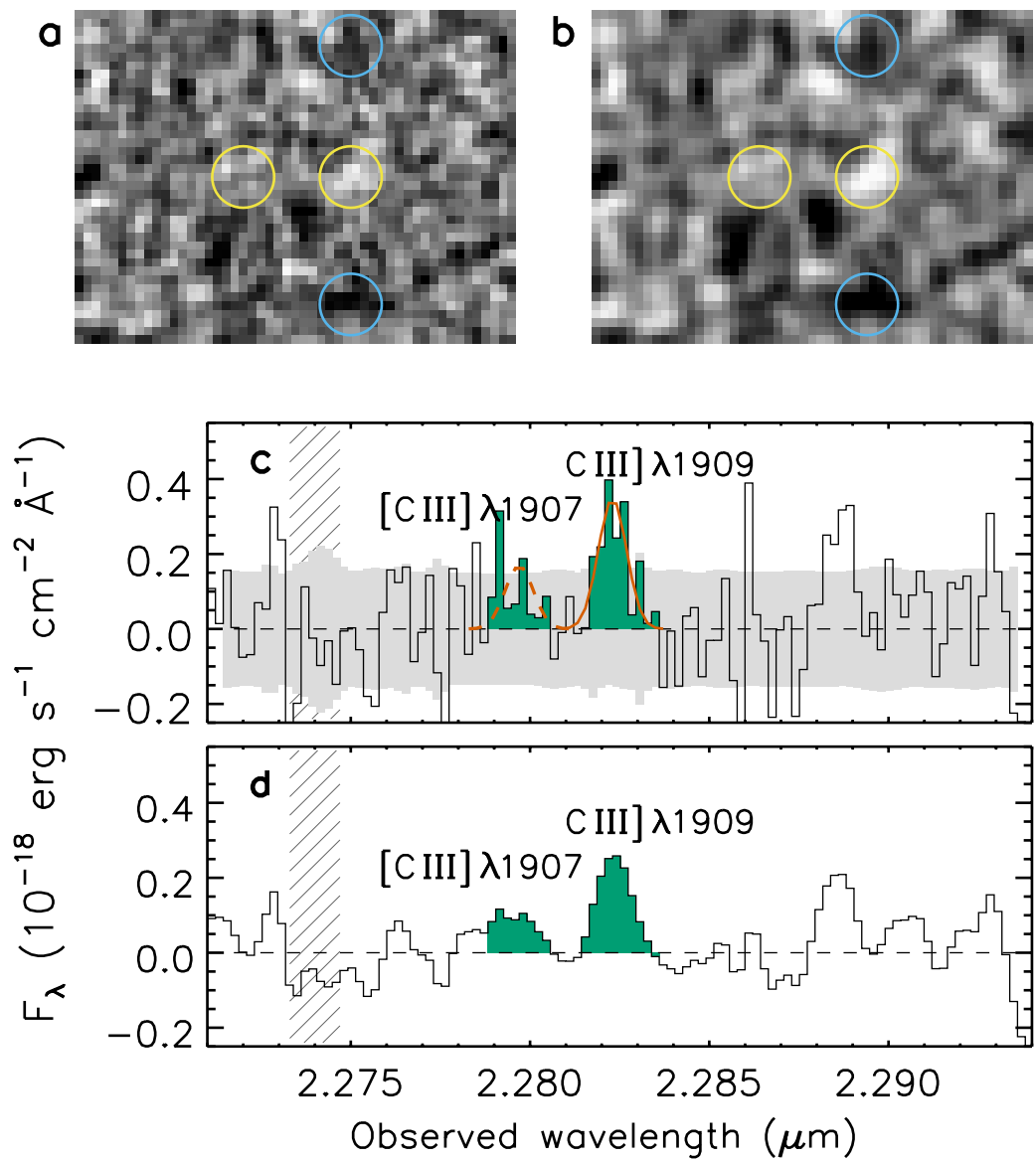

**Extended Data Fig. 1 Detection of the [C III] λ1907 and C III] λ1909 emission lines.**
Same as Fig. 2, but for the first half of the *K*-band data.

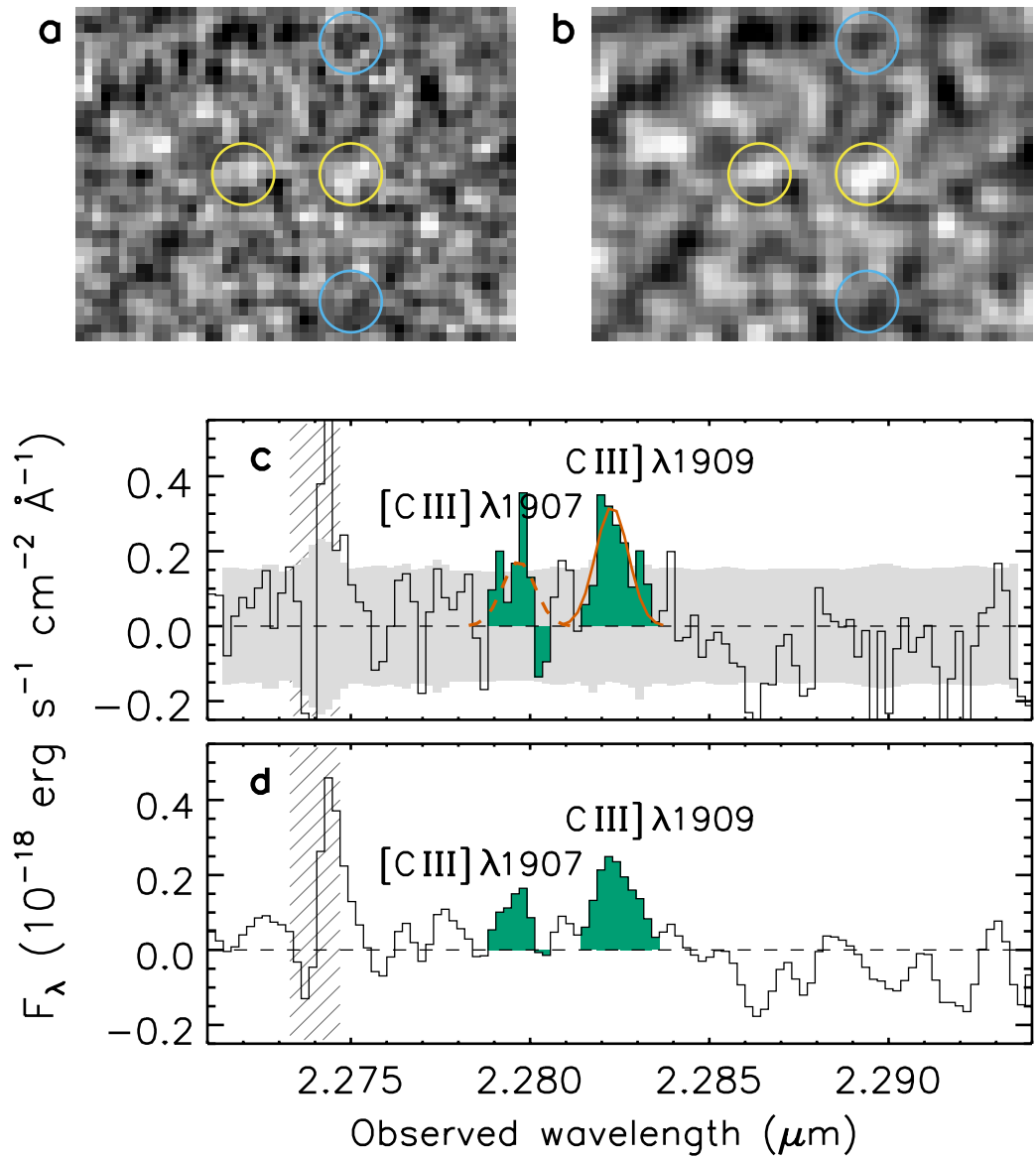

**Extended Data Fig. 2 Detection of the [C III] λ1907 and C III] λ1909 emission lines.**
Same as Fig. 2, but for the second half of the *K*-band data.

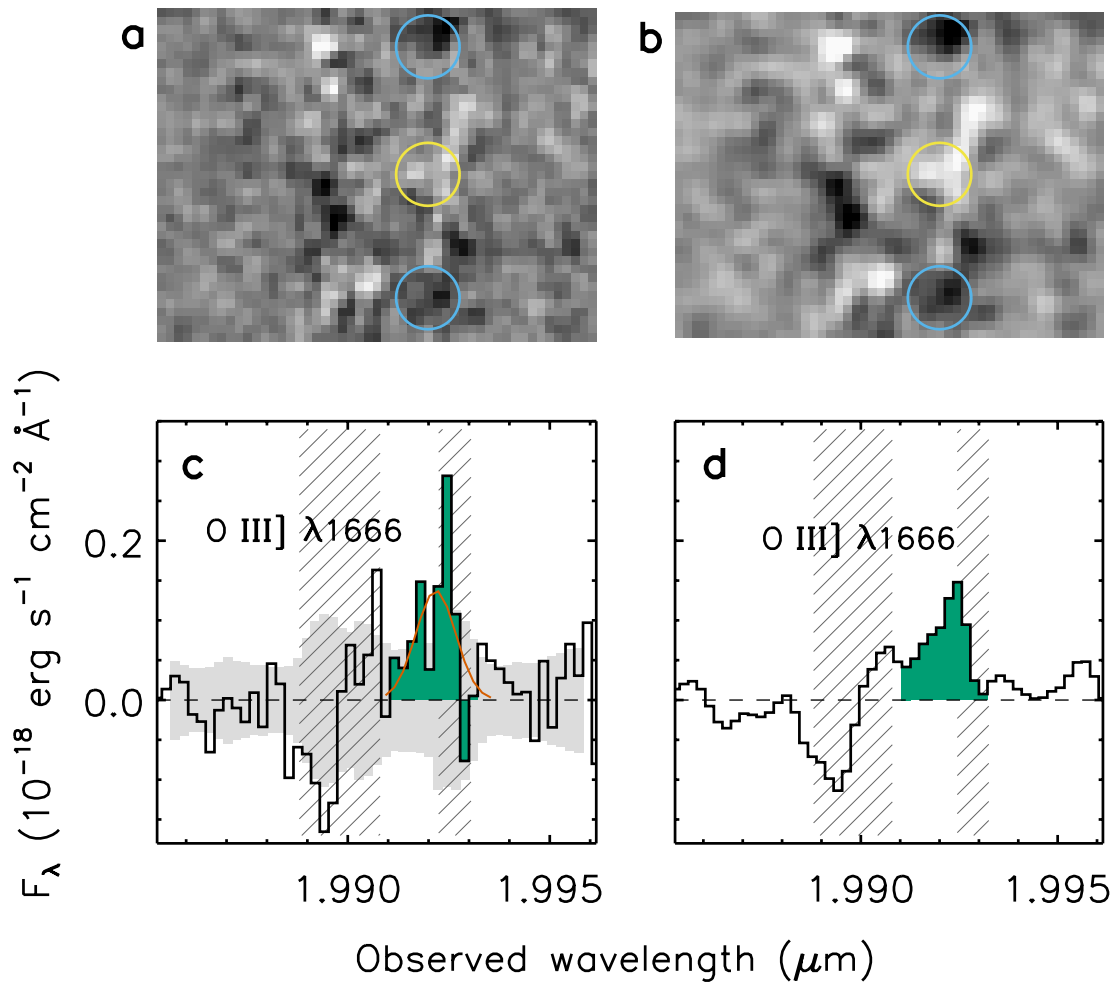

**Extended Data Fig. 3 Detection of the O III] λ1666 emission line.** (a): Part of the *K*-band 2D spectrum with the line (enclosed by the yellow circle) detected at 3.3σ significance. The two negative signals enclosed by the blue circles are the same pattern shown in Fig. 2. (b): A smoothed version of the 2D spectrum to better illustrate the line detection; a Gaussian filter with a size of 2.5 pixels is used. (c): Extracted 1D spectrum. The grey area represents the 1σ uncertainty region. The hatched areas indicate regions affected by OH skylines. The emission line is shown in green, with the vermillion solid profile being the best-fit Gaussian. (d): Same as (c), but from the smoothed 2D spectrum (b).

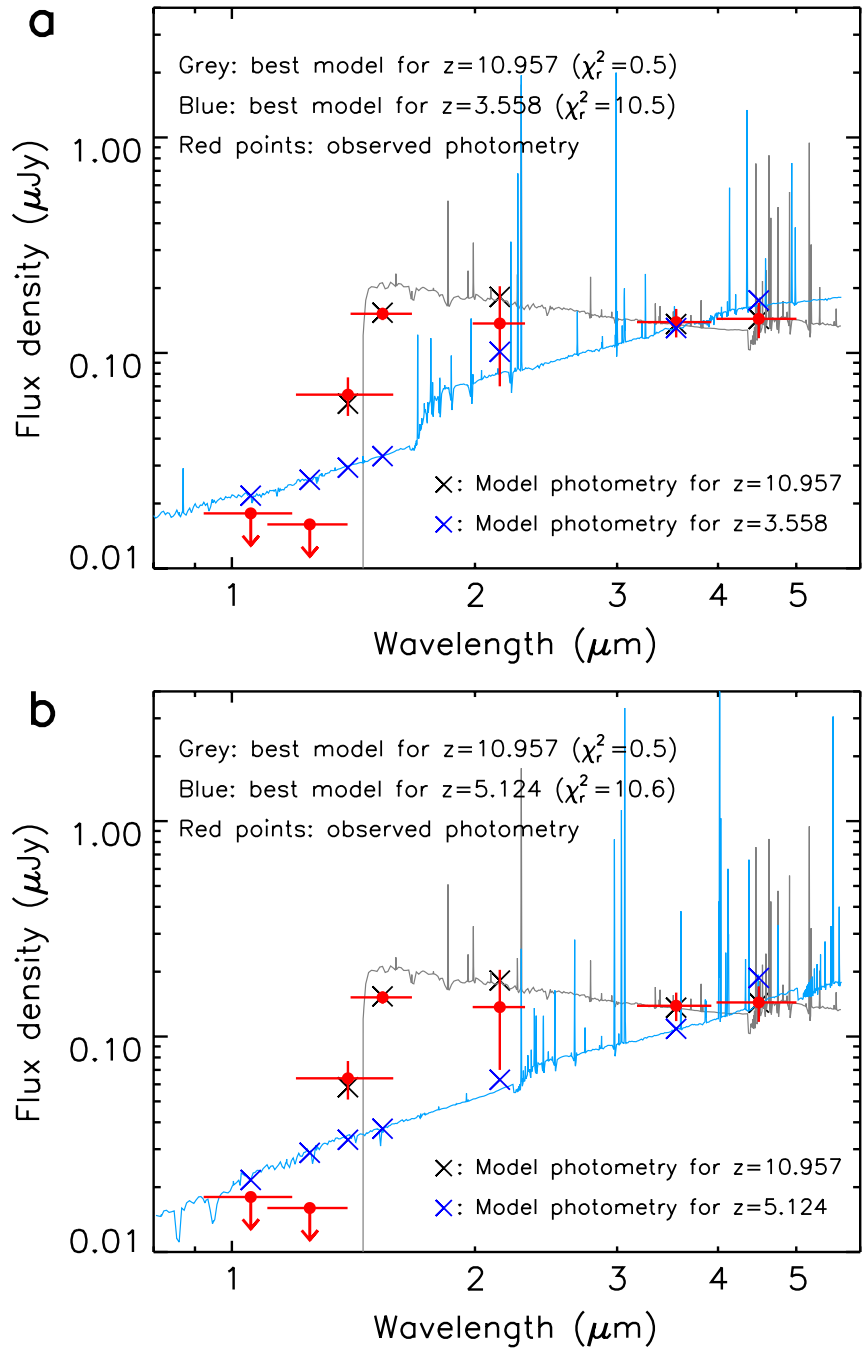

**Extended Data Fig. 4 SED modelling of GN-z11.** (a): SED modelling result using a fixed redshift $z = 3.558$, i.e., the emission line at 22823 Å is assumed to be [O III] $\lambda5007$. The red points with $1\sigma$ error bars are the observed photometric data points. The downward arrows indicate the $2\sigma$ detection upper limits. The horizontal errors indicate the wavelength ranges of the filters. The light blue spectrum represents the best model. The dark blue crosses represent the photometric points predicted by the model. They are inconsistent with the observed values. For comparison, the grey spectrum represents the best model using a fixed redshift $z = 10.957$. The model photometry (black crosses) is well consistent with the observed photometry. (b): Same as (a), but for a fixed redshift $z = 5.124$, i.e., the emission line at 22823 Å is assumed to be [O II] $\lambda3727$. The best model is also inconsistent with the observed photometry.